# Colloidal particles driven across periodic optical potential energy landscapes

Michael P. N. Juniper*,[1] Arthur V. Straube,[2] Dirk G. A. L. Aarts,[1] and Roel P. A. Dullens[1]

[1]Department of Chemistry, Physical and Theoretical Chemistry Laboratory,
University of Oxford, South Parks Road, OX1 3QZ Oxford, United Kingdom
[2]Department of Physics, Humboldt-Universität zu Berlin, Newtonstraße 15, 12489 Berlin, Germany

We study the motion of colloidal particles driven by a constant force over a periodic optical potential energy landscape. Firstly, the average particle velocity is found as a function of the driving velocity and the wavelength of the optical potential energy landscape. The relationship between average particle velocity and driving velocity is found to be well described by a theoretical model treating the landscape as sinusoidal, but only at small trap spacings. At larger trap spacings, a non-sinusoidal model for the landscape must be used. Subsequently, the critical velocity required for a particle to move across the landscape is determined as a function of the wavelength of the landscape. Finally, the velocity of a particle driven at a velocity far exceeding the critical driving velocity is examined. Both of these results are again well described by the two theoretical routes, for small and large trap spacings respectively. Brownian motion is found to have a significant effect on the critical driving velocity, but a negligible effect when the driving velocity is high.

## I. INTRODUCTION

The phenomenon of a particle travelling over a potential energy landscape is important to the behaviour of many physical systems of scientific interest and technological importance. This includes the diverse cases of counter-sliding rough surfaces [1], the movement of adatoms on atomic surfaces [2], and the motion of mobile rings on a poly-rotaxane [3]. Of particular current relevance are superconductor effects, such as DC driven Josephson junctions [4, 5] and charge density waves [6]. Such systems are, however, challenging to image [7], making microscopic scale motion difficult to study at easily accessible temperatures and pressures. Another well studied case is vortex motion in type-II superconductors [8–12]. Vortices may be directly imaged by techniques including Lorentz microscopy [13], Bitter decoration [14], or magneto-optical imaging [15], but direct, controllable access to microscopic motion is not available under readily accessible experimental conditions. Extensive work using computer simulation has been conducted [16–18], but there is still a requirement for model systems in which it is possible to examine behaviour in real space.

The experimental model system used in this paper is that of Brownian particles driven across a periodic optical potential energy landscape. Various techniques have been used to drive colloidal systems in optical potential energy landscapes [19–24], in order to address numerous problems, from tribology [19, 20] to particle sorting [21–24]. Of note is work considering the deflection of particles driven at an oblique angle across two-dimensional optical potential energy landscapes, where particle direction is dictated by the competition between the symmetry of the landscape and the direction of the driving force [22, 24, 25]. Furthermore, motion over both one- and two-dimensional potential energy landscapes has been used in models of friction, such as the Prandtl-Tomlinson model [26, 27] and the Frenkel-Kontorova model [27, 28].

The non-linearities in systems driven far from equilibrium by an external force have garnered recent interest in theory and experiment [29–33]. Considerable attention has been given to the problem of colloidal particles diffusing in a periodic potential [34–36], and diffusing over threshold potentials [37]. Further to this, the behaviour in a *tilted* periodic potential has been examined, with the bias leading to transport effects [38–44], giant [45, 46] or suppressed [47] diffusion.

In this article, we particularly focus on the critical driving velocity, upon changing the optical potential energy landscape from sinusoidal to non-sinusoidal by tuning the spacing between the optical traps constituting the landscape. We also study the average particle velocity well above the critical velocity. We compare our experimental results to a simple theoretical framework that describes the potential energy landscape in the limit of small and large trap spacing. We show that the Brownian motion of the particles need only be taken into account close to the critical driving velocity.

The paper is organised as follows. In Sec. II, we establish a simple theoretical model to explain the landscape dependent dynamics of the driven particles. The experimental methods are outlined in Sec. III, and their results are presented and compared to the theory in Sec. IV. Finally, we present our conclusions in Sec. V.

## II. THEORY

The following Langevin equation is used to describe the overdamped motion of a spherical Brownian particle driven by a constant force across a (periodic) potential

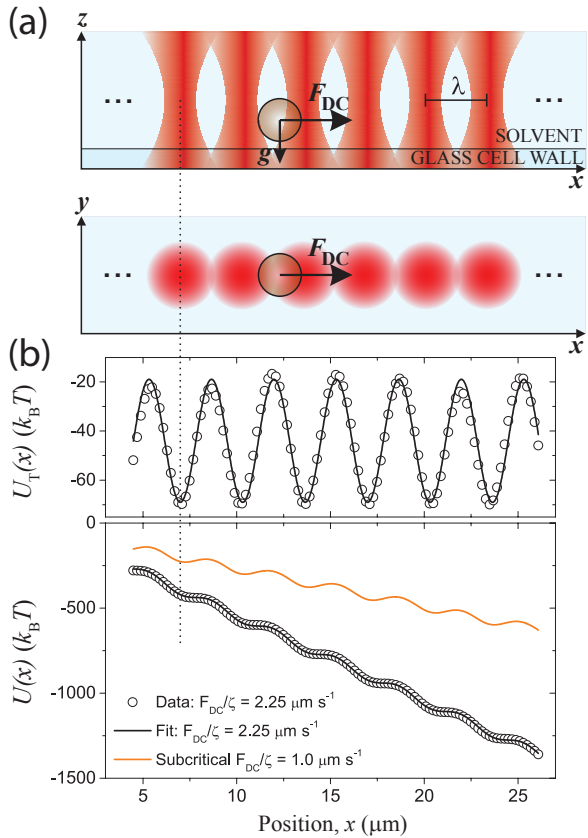

FIG. 1: (a) Schematic of the experimental geometry. A one dimensional periodic optical potential energy landscape, $U_T(x)$, is generated by a line of overlapping optical traps created from AOD-timeshared focused laser spots separated by a spacing $\lambda$. A spherical colloidal particle sedimented to the bottom of the sample cell is driven across the landscape by a constant force $F_{DC}$. (b) The optical potential energy landscape, $U_T(x)$, corresponding to a trap spacing of $\lambda = 3.5\mu$m and a laser power per trap of 0.75 mW. The tilted 'washboard' potential, $U(x)$ for $F_{DC}/\zeta = 2.25$ µm s$^{-1}$ is shown in the lower panel. The symbols are the experimental data and the solid black line a fit with sine function. The solid orange line in the lower panel illustrates a hypothetical tilted washboard potential corresponding to a subcritical driving force, which leads to finite barriers in the potential.

energy landscape, $U_T(x)$, (see Fig. 1) [34, 48]:

$$\zeta \frac{dx(t)}{dt} = F_{DC} + F_T(x) + \xi(t), \qquad (1)$$

where the instantaneous particle velocity, $v(x,t) = dx/dt$, at position $x$ and time $t$, depends on the constant (DC) driving force, $F_{DC}$, the force from the optical potential energy landscape, $F_T(x) = -\partial U_T/\partial x$, the Brownian force, $\xi(t)$, and the friction coefficient, $\zeta$. The impact of thermal fluctuations is modeled by Gaussian white noise, such that $\langle \xi(t) \rangle = 0$ and $\langle \xi(t)\xi(t') \rangle = 2\zeta k_B T \delta(t-t')$, where $k_B T$ is thermal energy. Thus, in the case $U_T(x)$ represents a spatially periodic landscape with a wavelength, $\lambda$, Eq. (1) describes the motion of a Brownian par-ticle in a tilted 'washboard' potential, $U(x) = -x F_{DC} + U_T(x)$, where $F_{DC}$ determines the tilt (see Fig. 1).

The relative importance of the deterministic and stochastic parts of Eq. (1) may be quantified using the Péclet number. We define this in our context as the ratio of the time taken for the particle to diffuse over a distance equivalent to one wavelength of the landscape (the 'Brownian time', $\tau_B = \lambda^2/D$ with $D = k_B T/\zeta$ being the diffusion coefficient) and the time taken for the particle to be driven over one wavelength of the landscape, $\tau_D = \lambda/\overline{v}$, where $\overline{v}$ is the average particle velocity:

$$\text{Pe} = \frac{\tau_B}{\tau_D} = \frac{\zeta \lambda \overline{v}}{k_B T}. \qquad (2)$$

The time taken for the particle to be driven over one wavelength of the landscape, $\tau_D$, results from the balance between the driving force and the force due to the optical potential energy landscape, and thus contains the average particle velocity, rather than the driving velocity.

When Pe $\gg 1$, the effect of diffusion is negligible relative to the driving force, but as Pe $\to 1$, diffusion becomes more important. To simplify the analysis of Eq. (1), we will neglect the stochastic force term, $\xi(t)$, for now. This approximation is instructive and is justified because the Péclet number is much higher than unity for most driving velocities used here. The (deterministic) equation of motion thus becomes:

$$\zeta \frac{dx(t)}{dt} = F_{DC} + F_T(x). \qquad (3)$$

To define the periodic optical potential energy landscape, $U_T(x)$, we assume that the landscape extends infinitely, from trap $i = -\infty$ to trap $i = \infty$, with traps separated by a spacing $\lambda$. Each individual trap $i$ is modelled by a Gaussian well $V_i(x)$ of depth $V_0$ and stiffness $k$ [22, 24, 49, 50],

$$V_i(x) = -V_0 \exp\left[-\frac{k(x-\lambda i)^2}{2V_0}\right]. \qquad (4)$$

We stress that although in the vicinity of the trap centre, $|x - \lambda i| \ll \lambda$, Eq. (4) reduces to the conventionally used harmonic potential, $V_i(x) = k(x-\lambda i)^2/2$, the harmonic approximation generally fails to properly describe the energy landscape; see also Refs. [49, 51], where the non-harmonic nature of the optical potential is crucial for capturing equilibrium and non-equilibrium pattern formation. As shown in Ref. [50], individual potentials are additive, so the potential landscape may be expressed as $U_T(x) = \sum_{i=-\infty}^{\infty} V_i(x)$, which leads to an optical force

$$F_T(x) = -k \sum_{i=-\infty}^{\infty} (x-\lambda i) \exp\left[-\frac{k(x-\lambda i)^2}{2V_0}\right]. \qquad (5)$$

In the experiments, two main observables are considered: the average particle velocity over an integer number of wavelengths of the landscape, $\overline{v}$, and the critical driving velocity, $F_C/\zeta$, required for the particle to move.

Firstly we consider the average particle velocity. For the periodic landscape, the time, $\Delta t$, in which the particle passes a single wavelength of the landscape, $\lambda$, is (see Eq. (3)): $\Delta t = \zeta \int_{-\lambda/2}^{\lambda/2} [F_{DC} + F_T(x)]^{-1} \, dx$. It therefore follows that in the deterministic regime:

$$\bar{v} = \frac{\lambda}{\Delta t} = \lambda \left( \int_{-\lambda/2}^{\lambda/2} \frac{\zeta}{F_{DC} + F_T(x)} \, dx \right)^{-1}. \quad (6)$$

Next, the critical driving force required to cause the particle to overcome a maximum in the optical force is considered. By setting $dx/dt = 0$ in Eq. (3), we find a stationary solution, $x = x_0$, such that $F_{DC} + F_T(x) = 0$. This solution describes the locked state because the particle is pinned to the periodic landscape and shifted from one of its nearest local minima at $x_{min} = i\lambda$ ($i = 0, \pm 1, \pm 2, \dots$) by $\delta x$ such that $x_0 = x_{min} + \delta x$. The locked state exists only if the constant driving force is small enough, $F_{DC} < F_C$, that there are finite barriers in the full potential (see Fig. 1(b), orange line), with the critical force

$$F_C = \max_x \left[ -F_T(x) \right] = -F_T(x_*). \quad (7)$$

Here, $x_*$ is the position of the maximum in the optical force, defined by $F_T'(x_*) = 0$, where the prime denotes the derivative with respect to $x$. For $F_{DC} > F_C$ there exist no stationary solutions, $dx/dt \neq 0$. This regime corresponds to the sliding state, meaning that the particle is sliding across the landscape with a certain averaged speed. The transition from the locked to sliding state occurs when $F_{DC} = F_C$. With $F_T(x)$ given by Eq. (5), the critical force may be stated directly:

$$F_C = k \sum_{i=-\infty}^{\infty} (x_* - \lambda i) \exp\left[ -\frac{k(x_* - \lambda i)^2}{2V_0} \right]. \quad (8)$$

Note that in this regime $F_{DC}$ is greater than, but close to $F_C$, implying that the Péclet number is close to one and diffusion is important, as will be discussed further in Sec. II D. Equations (6) and (8) are rigorous, but not analytically tractable for the full optical landscape (Eq. (5)). We therefore make some approximations in the description of the optical potential energy landscape.

## A. Sinusoidal landscape: small trap spacing

Since the optical energy landscape is periodic, it may be expressed as a Fourier series:

$$U_T(x) = \frac{1}{2}a_0 + \sum_{m=1}^{\infty} \left[ a_m \cos\left( \frac{2\pi m x}{\lambda} \right) + b_m \sin\left( \frac{2\pi m x}{\lambda} \right) \right].$$

The calculation of the first Fourier coefficient, $a_m = (2/\lambda) \int_{-\lambda/2}^{\lambda/2} U_T(x) \cos(2\pi m x/\lambda) \, dx$, yields:

$$a_m = -\frac{2\sqrt{2\pi}V_0^{3/2}}{\lambda k^{1/2}} \exp\left( -\frac{2\pi^2 m^2 V_0}{\lambda^2 k} \right),$$

with $m = 0, 1, 2, \dots$. The second Fourier coefficient vanishes, $b_m \equiv 0$, because $U_T(x)\sin(2\pi m x/\lambda)$ is an odd function integrated within symmetric limits. As a result, the trapping potential $U_T(x)$ is represented as:

$$U_T(x) = -\frac{2\sqrt{2\pi}V_0^{3/2}}{\lambda k^{1/2}}$$
$$\times \left[ \frac{1}{2} + \sum_{m=1}^{\infty} \exp\left( -\frac{2\pi^2 m^2 V_0}{\lambda^2 k} \right) \cos\left( \frac{2\pi m x}{\lambda} \right) \right]. \quad (9)$$

In Ref. [50] it was demonstrated that if the trap spacing is sufficiently small, the velocity profile for a particle passing across the periodic potential is well described by a sinusoidal function. It is therefore asserted that for small $\lambda$, the potential may be approximated by the leading sinusoidal term. Indeed, as becomes evident from Eq. (9), at small $\lambda$ the amplitudes decay exponentially fast with $m$. For this reason, terms with $m > 1$ may be neglected in Eq. (9), and the optical trapping force follows directly:

$$F_T(x) \approx -\frac{4\sqrt{2}(\pi V_0)^{3/2}}{\lambda^2 k^{1/2}} \exp\left( -\frac{2\pi^2 V_0}{\lambda^2 k} \right) \sin\left( \frac{2\pi x}{\lambda} \right). \quad (10)$$

The critical driving force, $F_C$, is found from Eq. (7). Accordingly, solving the equation $F_T'(x_*) = 0$ for $F_T(x)$ as in Eq. (10) yields $x_* = \lambda/4 + i\lambda$ (where $i = 0, \pm 1, \pm 2, \dots$) so that the critical force for small trap spacing $\lambda$ becomes:

$$F_C = -F_T(x_*) = \frac{4\sqrt{2}(\pi V_0)^{3/2}}{\lambda^2 k^{1/2}} \exp\left( -\frac{2\pi^2 V_0}{\lambda^2 k} \right). \quad (11)$$

By taking into account Eqs. (10) and (11), the deterministic equation of motion, Eq. (3), is reduced to an Adler equation [52, 53]:

$$\zeta \frac{dx(t)}{dt} = F_{DC} - F_C \sin\left( \frac{2\pi x}{\lambda} \right), \quad (12)$$

which offers a much simpler solution to the average velocity, Eq. (6). The time taken for a particle to pass one wavelength of the landscape in this case is $\Delta t = \zeta \int_{-\lambda/2}^{\lambda/2} [F_{DC} - F_C \sin(2\pi x/\lambda)]^{-1} dx = \zeta\lambda(F_{DC}^2 - F_C^2)^{-1/2}$, leading to an average particle velocity of [54]:

$$\bar{v} = \begin{cases} 0, & \text{if } F_{DC} \leq F_C; \\ \frac{1}{\zeta}\sqrt{F_{DC}^2 - F_C^2}, & \text{if } F_{DC} > F_C. \end{cases} \quad (13)$$

These equations will be compared to experimental results for potential energy landscapes with small trap spacings.

## B. Nearest neighbour landscape: large trap spacing

Now the case of widely spaced traps is considered, by treating them as almost non-overlapping Gaussian traps (Eq. (4)). An approximation is made that the local optical potential may be described by one central trap, numbered for simplicity by $i = 0$, and its two nearest neighbours, $i = \pm 1$. Taking this approach, only terms with $|i| \leq 1$ are taken from the full sum for the optical potential energy landscape, i.e. $U_T(x) = \sum_{i=-1}^{1} V_i(x)$, leading to the following expression for the optical force:

$$F_T(x) = -k \sum_{i=-1}^{1} (x - \lambda i) \exp\left[-\frac{k(x - \lambda i)^2}{2V_0}\right]. \quad (14)$$

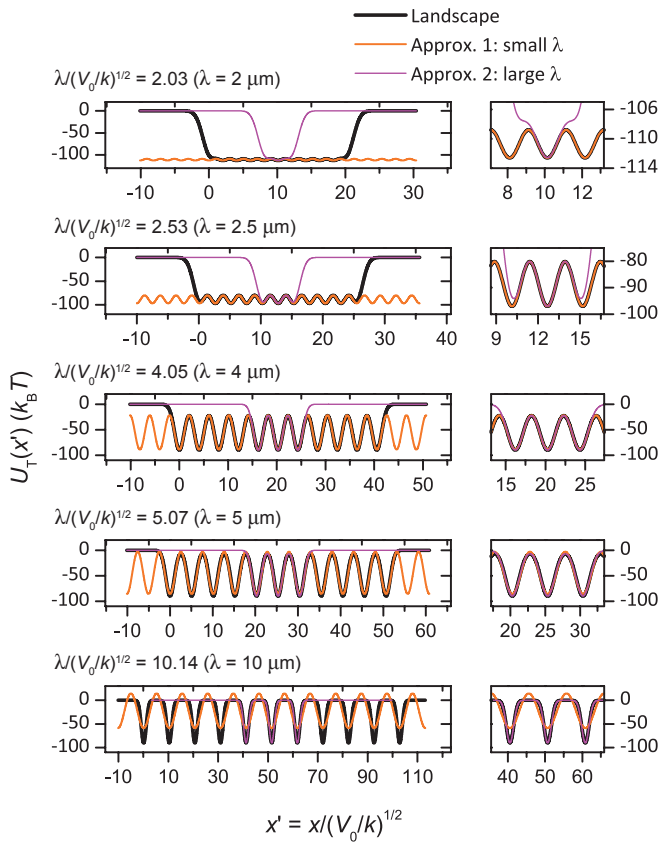

FIG. 2: Comparison of theoretical approximations for the optical potential energy landscape $U_T(x)$. — Periodic landscape composed of the sum of 11 individual Gaussian optical traps, at trap spacing $\lambda$ given in $\mu$m and units of $\sqrt{V_0/k}$, using typical trapping parameters $k = 3.8 \times 10^{-7}$ kg s$^{-2}$ and $V_0 = 90$ $k_B T$. — sinusoidal landscape: small $\lambda$, — nearest neighbour landscape: large $\lambda$.

This may be substituted into Eq. (6) to find the average particle velocity numerically.

In order to calculate the critical driving force for widely spaced traps, firstly the case is considered where the traps are infinitely spaced, $\lambda \to \infty$, and only a single term ($i = 0$) is taken from the sum in Eq. (14). The trapping force is then $F_T^\infty(x) = -kx \exp[-kx^2/2V_0]$, and maximising this as in Eq. (7) leads to the position of the maximum optical force, $x_*^\infty = \sqrt{V_0/k}$, which in turn provides an expression for the critical driving force of a single trap:

$$F_C^\infty = -F_T(x_*^\infty) = \sqrt{kV_0} \exp\left[-\frac{1}{2}\right].$$

Next, we take into consideration the optical potential of the nearest neighbouring traps ($i = \pm 1$) to account for the dependence of $F_C$ on the large but finite $\lambda$. Accordingly, by substituting an ansatz $x_* = x_*^\infty + \delta x_*$ into Eq. (14) and then solving $F_T'(x_*) = 0$, retaining only the leading term, we obtain the exponentially small correction $\delta x_* = -\sqrt{k/V_0}\lambda^2 \exp\left[-k\lambda^2/2V_0\right]$. The critical force for large $\lambda$ then follows from Eq. (7):

$$F_C = k \sum_{i=-1}^{1} (x_* - \lambda i) \exp\left[-\frac{k(x_* - \lambda i)^2}{2V_0}\right], \quad (15)$$

with

$$x_* = \sqrt{\frac{V_0}{k}} - \sqrt{\frac{k}{V_0}}\lambda^2 \exp\left[-\frac{k\lambda^2}{2V_0}\right]. \quad (16)$$

These expressions for the average velocity and the critical force will be compared to experimental measurements for periodic optical potential energy landscapes with large trap spacings.

## C. Comparing the approximations

Figure 2 shows periodic optical potential energy landscapes comprising eleven single Gaussian traps, with spacings varying from 2 $\mu$m to 10 $\mu$m, and typical trapping parameters of $k = 3.8 \times 10^{-7}$ kg s$^{-2}$ and $V_0 = 90$ $k_B T$. Note that we also give the trap spacings in units of $\sqrt{V_0/k}$, which is a measure for the width of a single trap. For comparison we additionally show the landscapes corresponding to the sinusoidal (small $\lambda$, as in Eq. (9) with a single term $m = 1$ only), and the nearest neighbour (large $\lambda$, as in $U_T(x) = \sum_{i=-1}^{1} V_i(x)$) approximations. The five large panels show the full landscape for each trap spacing, and the small panels show details of these, to highlight the comparisons between the different models.

Several features should be noted from the comparisons. Firstly, the height of the barrier in the landscape increases with trap spacing, until it is equal to the depth of a single trap, when the trap separation is very large ($\lambda = 10$ $\mu$m). At very small trap spacing ($\lambda = 2$ $\mu$m), the barrier is on the order of 4 $k_B T$, easily accessible by diffusion alone. However already at $\lambda = 2.5$ $\mu$m, the barrier is 20 $k_B T$, making diffusion from one minimum to another unlikely. Secondly, it is observed that the sinusoidal (small $\lambda$) approximation describes the full landscape very well for 2 $\mu$m $\leq \lambda \leq 4.5$ $\mu$m. Above this, however, both the form and depth of the potential landscape are poorly described. Conversely, the nearest neighbour (large $\lambda$) approximation does not describe the landscape well below $\lambda = 4$ $\mu$m, but is a very good description when $\lambda \geq 4$ $\mu$m. The small and large $\lambda$ approximations therefore cover the whole interval of required values of $\lambda$ and have a small overlapping interval, 4 $\mu$m $\leq \lambda \leq 4.5$ $\mu$m.

### D. The effect of Brownian noise

So far, a deterministic Langevin equation has been used, which is only valid at high Péclet number, Pe. Just above the critical driving velocity, the particle velocity is very low and Pe ∼ 1, so that the stochastic force term, $\xi(t)$, is of the same magnitude as the driving force, $F_{DC}$. In this regime, it is therefore necessary to consider the effect of Brownian motion on the particle velocity.

The average velocity of an overdamped Brownian particle in a tilted periodic potential, $U(x) = -xF_{DC} + U_T(x)$, can be expressed as [34, 54, 55]

$$\overline{v} = \frac{\lambda k_B T}{\zeta \mathcal{J}} \left( 1 - \exp\left[ -\frac{F_{DC}\lambda}{k_B T} \right] \right) \tag{17}$$

with

$$\mathcal{J} = \int_{-\lambda/2}^{\lambda/2} \exp\left[ -\frac{U(x)}{k_B T} \right] \mathrm{d}x \int_x^{x+\lambda} \exp\left[ \frac{U(x')}{k_B T} \right] \mathrm{d}x'. \tag{18}$$

Generally, for an arbitrary periodic potential $U_T(x)$, integral (18) admits no analytic representation and Eq. (17) has to be computed numerically. For the case of small $\lambda$, however, the optical potential is sinusoidal (see Sec. II A) and Eq. (18) can be explicitly integrated to yield $\mathcal{J} = \lambda^2 \exp(-\pi\mathcal{F}) |I_{i\mathcal{F}}(\mathcal{F}_C)|^2$ [55]. Here, $I_{i\mathcal{F}}(x)$ is the modified Bessel function of the first kind, and

$$\mathcal{F} = \frac{F_{DC}\lambda}{2\pi k_B T}; \quad \mathcal{F}_C = \frac{F_C\lambda}{2\pi k_B T},$$

with $F_C$ taken from Eq. (11). Thus, the average velocity of a particle driven over a landscape with small $\lambda$ is:

$$\overline{v} = \left( \frac{2k_B T}{\lambda \zeta} \right) \frac{\sinh(\pi\mathcal{F})}{|I_{i\mathcal{F}}(\mathcal{F}_C)|^2}. \tag{19}$$

Critical driving forces are determined numerically from these expressions, as described in Sec. III D below.

## III. EXPERIMENTAL METHODS

### A. Colloidal model system

The colloidal system used is composed of Dynabeads M-270 (diameter 3 μm), in 20% EtOH$_{aq}$, held in a quartz glass cell (Hellma) with internal dimensions of 9 mm × 20 mm × 200 μm. After filling we wait for sufficient time for residual flows to be absent. The particles are much more dense than the solvent, so they sediment into a single layer near the lower wall of the glass sample cell. The friction coefficient, $\zeta$, in the absence of any optical landscape is found from diffusion to be $\zeta = 9.19 \times 10^{-8}$ kg s$^{-1}$, which is slightly higher than would be expected from Stokes friction alone ($\zeta_{Stokes} = 6\pi\eta a$), due to the proximity of the wall. Particle concentration is low, so that only a single particle is visible in the field of view.

### B. Experimental setup and parameters

The experimental setup consists of an infra-red (1064 nm) laser, controlled using a pair of perpendicular acousto-optical-deflectors (AODs), and focused using a 50×, NA = 0.55 microscope objective [50]. One-dimensional periodic optical landscapes, with trap spacings 2.5 μm ≤ $\lambda$ ≤ 10 μm, are generated in Aresis Tweez software, controlled using a LabView interface. The traps are time-shared at 5 kHz, such that on the timescale of the particles (with a Brownian time of ∼ 50 s, and a driven time of at least ∼ 1/3 s per trap spacing), the traps form a constant potential energy landscape.

The laser power and the total number of traps are held constant throughout the experiments, so that the laser power per trap (and hence the trapping parameters $k$ and $V_0$) are consistent. A laser power of 350 mW is set, and 46 traps are used, corresponding to ∼ 0.75 mW per trap at the sample position. This gives typical trapping parameters for trap stiffness, $k = 3.8 \times 10^{-7}$ kg s$^{-2}$, and trap strength, $V_0 = 90 \, k_B T$. The particle is forced closer to the wall by the optical potential energy landscape, thus increasing the friction over the value quoted above. In order to account for this difference, we measured the friction coefficient felt by a particle in a single trap with the parameters described here, and found a value of $\zeta = 1.07 \times 10^{-7}$ kg s$^{-1}$, a difference of around 10% relative to the case where the optical landscape is absent. As this variation is much less than the variation in the velocity due to the optical landscape, this effect does not significantly affect our measurements. The number of traps which fit in the field of view changes with trap spacing, so excess traps are positioned at the edges of the field of view, in lines parallel to the experimental landscape. These extra traps are sufficiently far away as to not influence the experiment, and have the added advantage of catching extra particles which diffuse into the field of view.

The driving force is provided by a PI-542.2CD piezo-stage, controlled using the LabView interface, moving at 0.05 μm s$^{-1}$ ≤ $F_{DC}/\zeta$ ≤ 10 μm s$^{-1}$.

### C. Average velocity experiments

Images are focused onto a Ximea CMOS camera using a 40×, NA = 0.50 microscope objective. Experimental parameters are set automatically in the LabView interface, and six repeats are made at each driving velocity for each trap spacing. Particle position is recorded live at 40 Hz from the camera image. Average velocity is found by linearly fitting to a graph of $x(t)$, over an integer number of wavelengths of the landscape. Instantaneous velocity is found as the numerical derivative of the $x(t)$ data. The measured average properties have typically been averaged over six repeats, and the error bars correspond to the standard deviation of the repeats.

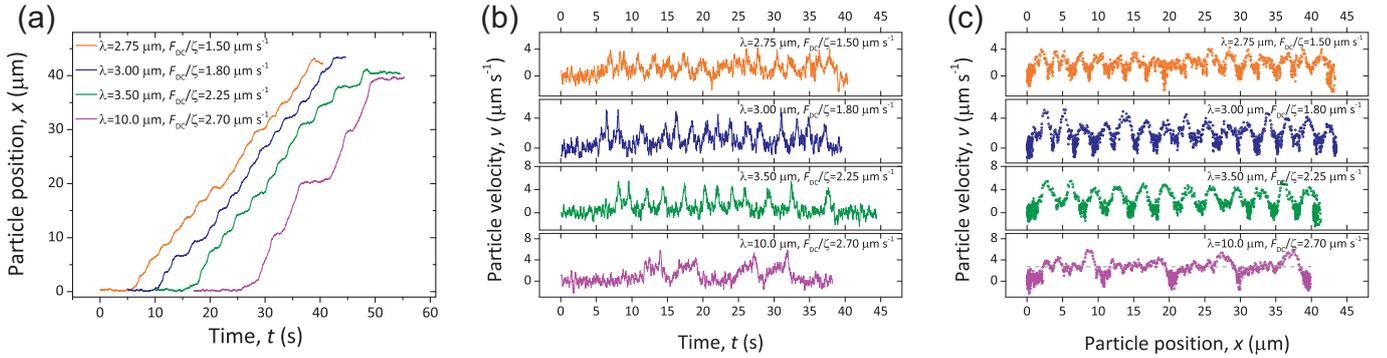

FIG. 3: Trajectories of Brownian particles driven over a periodic potential. (a) Individual particle trajectories for cases with differing trap spacings and driving velocities, but the same average velocity. Lines are spaced in $t$ for ease of comparison. (b) Particle velocity as a function of time for the four trajectories in (a). (c) Particle velocity as a function of position, for the four trajectories in (a). The dashed line in the right panel is at $v = 2.7$ µm s$^{-1}$

### D. Critical velocity experiments

The driving velocity is iterated to find the critical driving velocity at which the particle starts to slide across the optical potential energy landscape, with a maximum resolution of 0.05 µm s$^{-1}$, which constitutes the experimental uncertainty. A particle is said to be pinned if it does not move after the stage has moved 100 µm, or three minutes has elapsed, whichever happens first.

To determine the critical driving velocity from Eqs. (17) (large $\lambda$) and (19) (small $\lambda$), a numerical search is conducted to find the driving velocity at which the particle average velocity first exceeds a cut-off, set to the maximum resolution of the experiments (0.05 µm s$^{-1}$). Thus, at each trap spacing (with increments of $\Delta\lambda = 0.1$ µm), the driving velocity is increased until $\overline{v}$ is found to exceed 0.05 µm s$^{-1}$, and that driving velocity is then defined as the critical driving velocity. The results of these numerical experiments are presented and compared to experimental results in Sec. IV.

### IV. RESULTS AND DISCUSSION

Results are presented from experiments in which a colloidal particle was driven across a periodic optical potential energy landscape. Figure 3(a) shows four particle trajectories with the same average velocity, but each driven with different driving velocities, $F_{DC}/\zeta$, across a periodic optical potential energy landscape with a different wavelength, $\lambda$. The plateaus are spaced by the trap spacing as they are caused by the particle sitting in a trap for a period of time. The distribution of the waiting times between 'hops' from one trap to the next is a result of the combined effect of the external drive and thermal fluctuations, and does not reflect the uniformity of the underlying optical potential energy landscape.

Figure 3(b) shows the particle velocities, $v(t)$, obtained from the four trajectories in Fig. 3(a). When the particle is delayed in the vicinity of a trap centre, its velocity is close to zero. When a particle hops to the next minimum in the optical landscape, there is a spike in the velocity. The irregularity with which the particle hops to another trap illustrates the importance of thermal fluctuations at relatively low driving velocities.

Combining the $y$-axes of Figs. 3(a) and 3(b) leads to a measure of velocity as a function of position. Figure 3(c) shows $v(x)$ for the four cases shown in the previous two panels. The regions of minimal velocity are now evenly spaced by $\lambda$, as expected from the periodic potential energy landscape, and the regions where the particle is almost stationary are roughly the same size. A numerical integration of $v(x)$, which is directly proportional to $F_T(x)$, indeed yields a uniform and periodic $U_T(x)$, as shown for $\lambda = 3.5$ µm in Fig. 1(b). In the case of $\lambda = 10$ µm, when the traps are widely spaced, the particles regain the driving velocity (dashed line) after they have escaped the influence of each individual trap.

### A. Average particle velocity

The first main observable from the experiments is the average velocity of a particle travelling over many wavelengths of the periodic optical potential energy landscape. Figure 4 shows the dependence of the average particle velocity, $\overline{v}$, on the driving velocity, $F_{DC}/\zeta$, at a trap spacing of $\lambda = 3.5$ µm. At low driving velocities, the particle does not move across the landscape. The driving force then reaches a critical value, $F_{DC} = F_C$, after which the average particle velocity rises sharply from zero, before the gradient decreases, and ends roughly collinear to the line for the case of no traps. The average velocity will always be lower than the case of no traps, due to the time the particle is delayed by each trap (see Fig. 3). The experimental data are fitted with the deterministic expression for the average velocity for a particle driven over a sinusoidal potential energy landscape, $\zeta\overline{v} = \sqrt{F_{DC}^2 - F_C^2}$, see Eq. (13), which describes the data well, and gives a critical driving velocity of 1.8 µm s$^{-1}$, which is very similar to its direct measurement using the approach described in Sec. III D.

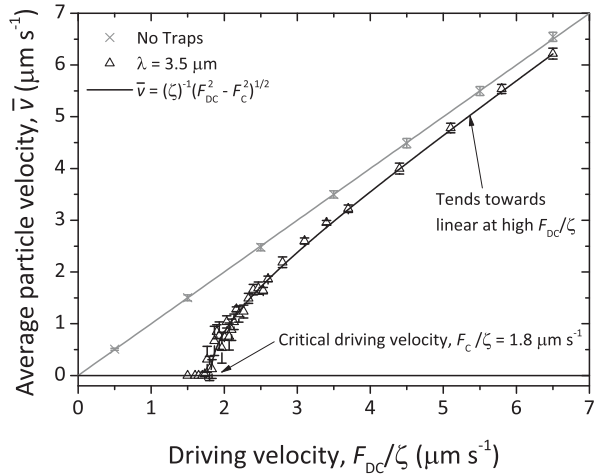

FIG. 4: Average particle velocity against driving velocity for $\lambda = 3.5$ μm, compared to the case of no traps. △ experimental data (error bars: standard deviation of repeats). — fit corresponding to a sinusoidal optical potential energy landscape, $\zeta\overline{v} = \sqrt{F_{DC}^2 - F_C^2}$ with fitting parameter $F_C$. -×- no traps.

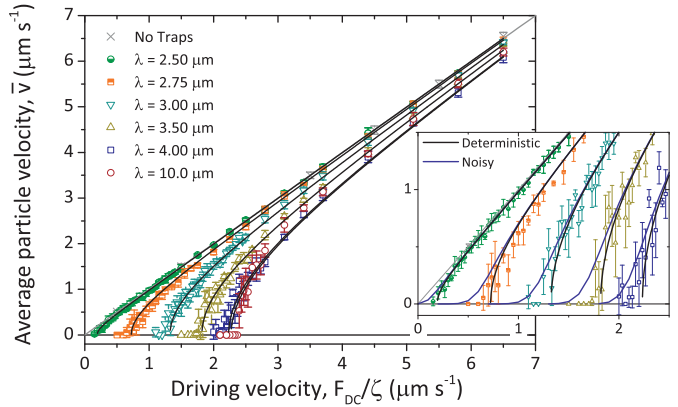

FIG. 5: Average particle velocity against driving velocity for varying trap spacing. ⊖, □, ▽, △, □, ○ experimental data (error bars: standard deviation of repeats). — fit corresponding to a sinusoidal optical potential energy landscape, $\zeta\overline{v} = \sqrt{F_{DC}^2 - F_C^2}$. -×- no traps. Inset: the effect of Brownian noise on particle velocity close to the critical driving velocity. — numerical solutions for a Brownian particle driven over a sinusoidal optical potential energy landscape, Eq. (19).

Next, the dependence of the average particle velocity on the driving velocity for different trap spacing is considered. Figure 5 shows the mean particle velocity as a function of driving velocity for six trap spacings, from $\lambda = 2.5$ μm to $\lambda = 10$ μm. All of the experimental lines show the same shape, but remain below the no trap line. The critical driving velocity increases with $\lambda$, which is discussed further in Sec. IV B. Each data set is fitted with Eq. (13), which describes the average velocity of a particle driven over a sinusoidal optical potential energy landscape in the absence of noise. This fit is accurate up to $\lambda = 4$ μm, but fits less well at $\lambda = 10$ μm. This is expected, as the sinusoidal nature of the potential energy landscape only holds for small trap spacing (see Fig. 2). At higher driving velocity, there is a decrease in average velocity as trap spacing increases. This is due to the time the particle is delayed by each trap, which increases with trap spacing due to the increase in the critical force, until a plateau at around $\lambda = 4$ μm, when the traps no longer overlap (see Fig. 7, described in Sec. IV C).

The behaviour at low driving velocity is less well described by the deterministic equation. The inset in Fig. 5 shows the lower parts of the curves for $\lambda = 2.5$ μm to $\lambda = 4$ μm, with the previously described fits and numerical solutions to the stochastic Langevin equation, Eq. (19), for a sinusoidal optical potential energy landscape. The effect of Brownian noise is only noticeable at very low average velocity, when Pe ~ 1. This manifests as a small deviation from the deterministic velocity While the deterministic velocity abruptly goes to zero at the critical driving velocity, the fits to Eq. (19) show a clear 'smoothing' of the average velocity upon approaching the critical driving velocity, in agreement with the experimental data.

## B. Critical driving velocity

Now the dependence of the critical driving velocity on the trap spacing is considered in more detail, and compared to theoretical predictions excluding and including Brownian noise. The experimental data shown in Figure 6 show the critical driving velocity, obtained as described in Sec. III D, as a function of the trap spacing. It is observed that at small trap spacings there is essentially no critical driving velocity, as the height of the barriers between the minima in the periodic optical potential energy landscape are on the order of a few $k_BT$, and the particle is able to diffuse across the landscape (see Fig. 2). At $\lambda \approx 2$ μm, the critical driving velocity increases sharply, until it plateaus at $\lambda \approx 5$ μm, with a value of $F_C/\zeta = 2.3$ μm s$^{-1}$, when the critical driving force becomes the force required to escape a single isolated Gaussian trap.

Four theoretical predictions are plotted on Fig. 6. Firstly, the dotted line and the dashed line show the deterministic solutions for the small $\lambda$ case, Eq. (11), and the large $\lambda$ case, Eq. (15), respectively. The deterministic expression for a sinusoidal landscape (small $\lambda$) is effective from $\lambda = 0$ to $\lambda \approx 4$ μm, while the deterministic expression for the nearest neighbour landscape (large $\lambda$) works better for large trap spacings. These deterministic theoretical predictions generally follow the same trend as the experimental data, though the critical driving velocity increases from zero at too low a trap spacing, and the plateau at large $\lambda$ lies at too high a value of $F_C/\zeta$. This is because around the critical driving velocity the Péclet number is on the order of unity, and the experimental

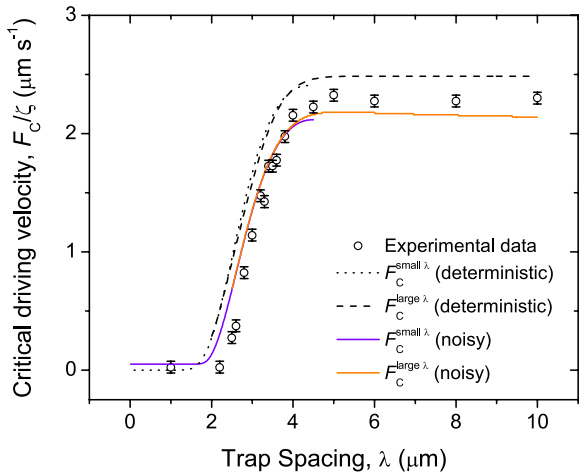

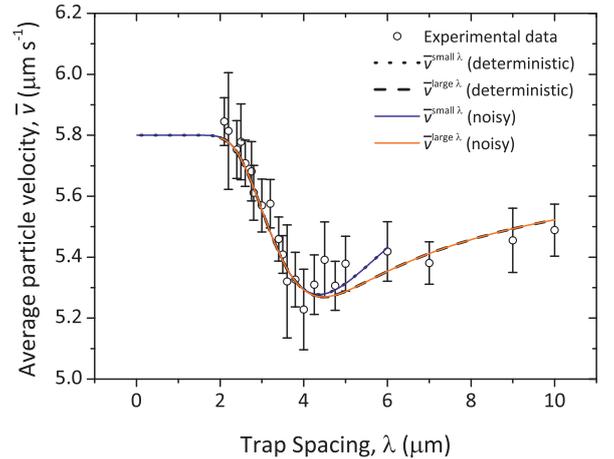

FIG. 6: Critical driving velocity as a function of trap spacing. ○ experimental data (error bars: experimental uncertainty). Lines: theoretical predictions. Sinusoidal potential energy landscape: ······ deterministic force, Eq. (11); —— stochastic force, according to Eq. (19). Nearest neighbour landscape: - - - deterministic force, Eq. (15), —— stochastic force, according to Eq. (17).

FIG. 7: Effect of the trap spacing on the average velocity of a particle at a driving velocity of 5.8 μm s⁻¹. ○ experimental data (error bars: standard deviation of repeats). Sinusoidal potential energy landscape: ··· deterministic force, Eq. (13), —— stochastic force, Eq. (19). Nearest neighbour landscape: - - - deterministic force, Eq. (6), —— stochastic force, Eq. (17).

$F_C/\zeta$ will be lowered due to the effect of Brownian noise, hence the deterministic predictions will overestimate the experimental critical velocity.

The numerical solutions including Brownian noise for the small $\lambda$ case (Eq. (19)) and the large $\lambda$ case (Eq. (17)) are also shown in Fig. 6. These predictions show the same trends as the deterministic lines, but have generally lower values. The numerical results including noise compare very favourably with the experimental results, with two caveats resulting from the fact that the meaning of the experimental and numerical critical velocities is not identical (see Sec. III D). Firstly, whereas the experimentally measured critical driving velocity is that required for a particle to escape a single minimum in the landscape within a reasonable period of time, the numerical method gives a critical driving velocity equal to the DC driving velocity required to achieve a certain threshold average velocity (set to 0.05 μm s⁻¹, see Sec. III D). Therefore, at very small trap spacings, where the barriers of the landscape are on the order of $k_B T$, the presence of noise negates these barriers and the average velocity is approximately equal to the applied driving velocity. The second caveat applies at large $\lambda$, where a small decrease in the numerically determined critical velocity is observed with increasing $\lambda$ as the impact of the traps on the average velocity becomes less pronounced compared to the space between them.

### C. Driving far above the critical driving velocity

When the particle is driven by a velocity well in excess of the critical driving velocity, the effect of the trap spacing on the average particle velocity is a result of the amount of time the particle is delayed by each trap (see Fig. 5). Figure 7 shows this effect more clearly by plotting the average particle velocity against the trap spacing, for $F_{DC} \gg F_C$. The experimental data show that velocity markedly decreases between $\lambda$ = 2 μm and $\lambda$ = 4 μm, before gradually increasing again.

Four theoretical predictions are plotted on Fig. 7. The deterministic solutions for the small $\lambda$ case, Eq. (13) and the large $\lambda$ case, Eq. (6), are shown along with the noisy solutions for small $\lambda$, Eq. (19), and large $\lambda$, Eq. (17), with $F_T(x)$ from Eq. (14). Here there is no difference between the predictions from the deterministic and noisy cases, which is consistent with the driving velocities in this regime being far beyond the critical velocity, so that the Péclet number is large and Brownian noise is thus unimportant. Again the small $\lambda$ approximation works up to $\lambda \approx 4$ μm, and the large $\lambda$ approximation works above $\lambda \approx 2.5$ μm, consistent with Fig. 6. The initial rapid decrease in $\bar{v}$ as $\lambda$ increases is due to the increased time the particle is delayed by each trap, due to the increased optical potential barrier. As the traps no longer overlap for larger $\lambda$, $\bar{v}$ increases, and the time spent *between* traps takes over, bringing the average velocity back up towards the driving velocity.

## V. CONCLUSIONS

We have studied the behaviour of Brownian particles driven by a constant force through quasi-one-dimensional periodic optical potential energy landscapes. Firstly, we have seen a critical driving velocity, below which the particle is pinned to the potential energy landscape. We have considered potential energy landscapes with a range of trap spacings, and the critical driving velocity has been found to depend on the wavelength. At small trap spacings, the critical driving velocity is very low, and it then increases, reaching a plateau when the individual traps are fully separated. Above the critical driving velocity the particle slides, with an average velocity that increases non-linearly with the driving velocity. The particle velocity is also found to depend on the landscape wavelength when the driving velocity is far in excess of the critical driving velocity. At small trap spacings, increasing the landscape wavelength reduces the average particle velocity, as each barrier in the optical potential becomes higher and delays the particle for longer. At larger trap spacings, average particle velocity increases again, as the particle velocity is determined by the time spent between fully separated traps.

We have made theoretical predictions corresponding to two different approximations for the optical potential energy landscape. When trap spacing is small the landscape is treated as sinusoidal, and when trap spacing is large it is treated as a sum of three individual Gaussian traps. These approximations have been broadened to include the effect of Brownian noise. The trend for the average particle velocity as a function of driving velocity and trap spacing has been shown to very accurately match the theoretical prediction from the small trap spacing approximation, up to a limiting landscape wavelength. Including the effect of Brownian noise allows more realistic prediction of the average particle velocity close to the critical driving velocity. Critical driving velocities themselves are also found to agree well with theoretical predictions, especially once the effect of Brownian noise has been taken into account. At higher driving forces, however, it is shown that the deterministic solutions alone provide an adequate description.

**Acknowledgements**

We thank Alice Thorneywork, Samantha Ivell and Urs Zimmermann for useful discussions. EPSRC is acknowledged for financial support.

[1] F. Heslot, T. Baumberger, B. Perrin, B. Caroli, and C. Caroli, Physical Review E **49**, 4973 (1994).

[2] P. Talkner, E. Hershkovitz, E. Pollak, and P. Hänggi, Surface Science **437**, 198 (1999).

[3] M. B. Pinson, E. M. Sevick, and D. R. M. Williams, Macromolecules **46**, 4191 (2013).

[4] B. D. Josephson, Physics Letters **1**, 251 (1962).

[5] P. W. Anderson and J. M. Rowell, Physical Review Letters **10**, 230 (1963).

[6] G. Grüner, Reviews of Modern Physics **60**, 1129 (1988).

[7] J. M. Carpinelli, H. H. Weitering, E. W. Plummer, and R. Stumpf, Nature **381**, 398 (1996).

[8] A. A. Abrikosov, Reviews of Modern Physics **76**, 975 (2004).

[9] R. Besseling, P. H. Kes, T. Drose, and V. M. Vinokur, New Journal of Physics **7**, 71 (2005).

[10] R. Besseling, Ph.D. thesis, Universiteit Leiden (2001).

[11] P. Martinoli, Physical Review B **17**, 1175 (1978).

[12] V. Gotcheva and S. Teitel, Physical Review Letters **86**, 2126 (2001).

[13] K. Harada, T. Matsuda, J. Bonevich, M. Igarashi, S. Kondo, G. Pozzi, U. Kawabe, and A. Tonomura, Nature **360**, 51 (1992).

[14] L. Y. Vinnikov, J. Karpinski, S. M. Kazakov, J. Jun, J. Anderegg, S. L. Budko, and P. C. Canfield, Physical Review B **67**, 092512 (2003).

[15] V. K. Vlasko-Vlasov, A. E. Koshelev, U. Welp, W. Kwok, A. Rydh, G. W. Crabtree, and K. Kadowaki, *Magneto-optical imaging of Josephson vortices in layered superconductors* (Springer, 2004), chap. Magneto-Optical Imaging, pp. 39–46.

[16] A. E. Koshelev and V. M. Vinokur, Physical Review Letters **73**, 3580 (1994).

[17] C. Reichhardt, C. J. Olson, and F. Nori, Physical Review Letters **78**, 2648 (1997).

[18] C. J. Olson, C. Reichhardt, and F. Nori, Physical Review Letters **81**, 3757 (1998).

[19] T. Bohlein, J. Mikhael, and C. Bechinger, Nature Materials **11**, 126 (2012).

[20] T. Bohlein and C. Bechinger, Physical Review Letters **109**, 058301 (2012).

[21] M. P. MacDonald, G. C. Spalding, and K. Dholakia, Nature **426**, 421 (2003).

[22] K. Ladavac, K. Kasza, and D. G. Grier, Physical Review E **70**, 010901 (2004).

[23] A. Jonas and P. Zemanek, Electrophoresis **29**, 4813 (2008).

[24] K. Xiao and D. G. Grier, Physical Review E **82**, 051407 (2010).

[25] M. Pelton, K. Ladavac, and D. G. Grier, Physical Review E **70**, 031108 (2004).

[26] V. L. Popov and J. A. T. Gray, Z. Angew. Math. Mech. **92**, 683 (2012).

[27] A. Vanossi and O. M. Braun, J. Phys.: Condens. Matter **19**, 305017 (2007).

[28] O. M. Braun and Y. S. Kivshar, *The Frenkel-Kontorova Model; Concepts, Methods, and Applications* (Springer, 2010).

[29] C. Reichhardt and C. J. Olson Reichhardt, Physical Review Letters **96**, 028301 (2006).

[30] V. Blickle, T. Speck, C. Lutz, U. Seifert, and C. Bechinger, Physical Review Letters **98**, 210601 (2007).

[31] A. I. Shushin, Journal of Physical Chemistry A **113**, 9065 (2009).

[32] J. M. Sancho and A. M. Lacasta, The European Physical Journal Special Topics **187**, 49 (2010).

[33] M. P. N. Juniper, A. V. Straube, R. Besseling, D. G. A. L. Aarts, and R. P. A. Dullens, Nature Communications **6**,

7187 (2015).

[34] P. Reimann, C. Van den Broeck, H. Linke, P. Hänggi, J. M. Rubi, and A. Pérez-Madrid, Physical Review E **65**, 031104 (2002).

[35] M. Šiler and P. Zemánek, New Journal of Physics **12**, 083001 (2010).

[36] C. Dalle-Ferrier, M. Kruger, R. D. L. Hanes, S. Walta, M. C. Jenkins, and S. U. Egelhaaf, Soft Matter **7**, 2064 (2011).

[37] G. Costantini and F. Marchesoni, Europhysics Letters **48**, 491 (1999).

[38] K. Lindenberg, A. M. Lacasta, J. M. Sancho, and A. H. Romero, New Journal of Physics **7**, 29 (2005).

[39] M. Evstigneev, O. Zvyagolskaya, S. Bleil, R. Eichhorn, C. Bechinger, and P. Reimann, Physical Review E **77**, 041107 (2008).

[40] Y. Roichman, V. Wong, and D. G. Grier, Physical Review E **75**, 011407 (2007).

[41] A. V. Straube and P. Tierno, Europhysics Letters **103**, 28001 (2013).

[42] X. G. Ma, P. Y. Lai, and P. G. Tong, Soft Matter **9**, 8826 (2013).

[43] X. G. Ma, P. Y. Lai, B. J. Ackerson, and P. G. Tong, Soft Matter **11**, 1182 (2015).

[44] X. G. Ma, P. Y. Lai, B. J. Ackerson, and P. G. Tong, Physical Review E **91**, 042306 (2015).

[45] S. H. Lee and D. G. Grier, Physical Review Letters **96**, 190601 (2006).

[46] P. Reimann, C. Van den Broeck, H. Linke, P. Hänggi, J. M. Rubi, and A. Perez-Madrid, Physical Review Letters **87**, 010602 (2001).

[47] S. Martens, A. V. Straube, G. Schmid, L. Schimansky-Geier, and P. Hänggi, Phys. Rev. Lett. **110**, 010601 (2013).

[48] P. Hänggi and F. Marchesoni, Rev. Mod. Phys. **81**, 387 (2009).

[49] A. V. Straube, A. A. Louis, J. Baumgartl, C. Bechinger, and R. P. A. Dullens, Europhysics Letters **94**, 48008 (2011).

[50] M. P. N. Juniper, R. Besseling, D. G. A. L. Aarts, and R. P. A. Dullens, Optics Express **27**, 28707 (2012).

[51] A. V. Straube, R. P. A. Dullens, L. Schimansky-Geier, and A. A. Louis, The Journal of Chemical Physics **139**, 134908 (2013).

[52] R. Adler, Proceedings of the IRE **34**, 351 (1946).

[53] R. E. Goldstein, M. Polin, and I. Tuval, Physical Review Letters **103**, 168103 (2009).

[54] M. Gitterman, *The Noisy Pendulum* (World Scientific Publishing, Singapore, 2008).

[55] R. L. Stratonovich, *Topics in the Theory of Random Noise*, vol. II (Gordon and Breach, New York, 1967).